\documentclass[prb,twocolumn,amssymb,showpacs]{revtex4}
\usepackage{graphicx}
\usepackage{amsmath}

\begin{document}

\title{Ferromagnetic - spin glass transition induced by pressure in the geometrically frustrated pyrochlore (Tb$_{1-x}$La$_x$)$_2$Mo$_2$O$_7$}
\author{A. Apetrei$^1$, I. Mirebeau$^1$, I. Goncharenko$^1$, D. Andreica$^{2,3}$ and P. Bonville$^4$
\\
}
\address{
$^1$Laboratoire L\'eon Brillouin, CEA-CNRS, CE-Saclay, 91191
Gif-sur-Yvette, France}
\address {$^2$Laboratory for Muon-Spin Spectroscopy, Paul Scherrer
Institut, 5232 Villigen-PSI, Switzerland}
\address {$^3$Babes-Bolyai
University, Faculty of Physics, 400084 Cluj-Napoca, Romania}
\address {$^4$Service de Physique de l'Etat Condens\'e,
CEA-CNRS, CE-Saclay,  91191 Gif-Sur-Yvette, France}

\begin{abstract}
We have studied (Tb$_{1-x}$La$_x$)$_2$Mo$_2$O$_7$ pyrochlores by
neutron diffraction and $\mu$SR at ambient and under applied
pressure. Substitution of Tb for La expands the lattice and induces
a change from a spin-glass like state ({\itshape x}=0) to a non
collinear ferromagnet ({\itshape x}=0.2). In the ferromagnetic
structure, the Tb moments orient close to their local anisotropy
axis as for an ordered spin ice, while the Mo moments orient close
to the net moment. The $\mu$SR dynamical relaxation rate shows a
cusp-like peak at the Curie temperature T$_C$ and a broad anomaly at
T$^\star$$<$T$_C$, suggesting a second transition of local or
dynamical nature. Under pressure, the long range order breaks down
and a spin glass-like state is recovered. The whole set of data
provide a microscopic picture of the spin correlations and
fluctuations in the region of the ferromagnetic-spin glass
threshold.

\end{abstract}

 \maketitle

In the pyrochlore compounds R$_2$M$_2$O$_7$, both rare earth
R$^{3+}$ and M$^{4+}$ transition metal form a three-dimensional
network of corner sharing tetrahedra. The pyrochlore lattice is
geometrically frustrated for antiferromagnetic (AF)
nearest-neighbour exchange interactions between Heisenberg spins. It
is also frustrated for ferromagnetic (F) interactions if spins are
constrained along their local anisotropy axes. 
Geometrical frustration can lead to intriguing short range ordered
phases with ground state degeneracy, such as spin liquids, spin
ices or chemically ordered spin glasses
\cite{Bramwell01,Greedan01}.

R$_2$Mo$_2$O$_7$ pyrochlores were extensively studied since their
electrical and magnetic properties exhibit an unusual variation
with the rare earth ionic radius r. Compounds with small ionic
radius R = Y, Dy and Tb show spin glass (SG) insulating behaviour,
whereas those with R = Gd, Sm and Nd show ferromagnetic (F)
metallic behaviour. (RR')$_2$Mo$_2$O$_7$ series with different
substitutions on the R$^{3+}$ site show a universal dependence of
the magnetic transition temperature on the average R$^{3+}$ ionic
radius\cite{Katsufuji00,Moritomo01}. It suggests that the sign of
Mo-Mo interactions controls the formation of the spin
glass$/$ferromagnetic state. Photoemission experiments and band
structure calculations\cite{Kang02,Solovyev03} point out that the
concomitant changes of the transport and magnetic properties come
from strong electron correlations in the Mo({\itshape t$_{2g}$})
band nearby the Fermi level.

Few microscopic studies of the magnetic order and spin fluctuations
were performed up to now in R$_2$Mo$_2$O$_7$. They all deal with
compounds far from the threshold ionic radius {\itshape r$_c$}
between SG and F regions. Y$_2$Mo$_2$O$_7$ and Tb$_2$Mo$_2$O$_7$
with {\itshape r}$<${\itshape r$_c$} were thoroughly
investigated\cite{Dunsiger96,Greedan91,Gaulin92,Gardner99}, since
the occurrence of a SG transition is surprising with regards to
their chemical order. On the F side, Nd$_2$Mo$_2$O$_7$ with
{\itshape r}$>${\itshape r$_c$} was also intensively studied
\cite{Yasui01,Taguchi01} due to its giant abnormal Hall effect. F
compounds nearby the threshold (R = Sm, Gd) are difficult to study
since they strongly absorb neutrons. The study of mixed
(RR')$_2$Mo$_2$O$_7$ compounds near the threshold is also intricate
since three magnetic ions are involved. Therefore neither the role
of R$^{3+}$ magnetism nor the evolution of the spin correlations and
fluctuations near the threshold have been clarified so far.

To understand the role of interatomic distances in the SG-F
transition, the most direct way is to combine applied pressure and
chemical pressure. This is done here in a single and "simple" system
(Tb$_{1-x}$La$_x$)$_2$Mo$_2$O$_7$, studied by neutron diffraction
and $\mu$SR at ambient and applied pressure. Substitution of Tb for
La expands the lattice, encompassing the threshold for a small La
content ({\itshape x}$\sim$0.06) which does not perturb much the Tb
lattice, since La is not magnetic. By neutron diffraction, we follow
the evolution of the spin correlation from SG to F state at a
microscopic level. For {\itshape x}=0.2 we observe a new ordered
magnetic structure, involving non collinear ferromagnetic
arrangement of both Tb- and Mo- moments. Under pressure the
{\itshape x}=0.2 sample transforms into a spin glass similar to
Tb$_2$Mo$_2$O$_7$, showing the equivalence of chemical and applied
pressures. While neutron diffraction probes magnetic correlations,
$\mu$SR measurements allows us to study spin fluctuations. For
{\itshape x}=0.2, we observe by $\mu$SR a second transition well
below the Curie transition. The two transitions seem to merge with
increasing pressure. The whole data provide a fully new picture of
the F-SG threshold.

The crystal structure of (Tb$_{1-x}$La$_x$)$_2$Mo$_2$O$_7$ powders
({\itshape x}=0, 0.05, 0.1, 0.15, 0.2) was characterized at 300 K by
combining X-ray and neutron diffraction. Rietveld refinements
performed with FULLPROF \cite{Carvajal93} show that all samples
crystallize in {\itshape Fd$\overline{3}$m} cubic space group, with
a lattice constant {\itshape a} between 10.312 $\AA$ ({\itshape
x}=0) and 10.378 $\AA$ ({\itshape x}=0.2). The {\itshape x}=0.10
sample ({\itshape a}= 10.346 $\AA$) is above the critical threshold
({\itshape a$_c$}= 10.332 $\AA$). The lattice constant was measured
under pressure for {\itshape x}=0 and 0.2 on the ID27 beam line of
the European Synchrotron Radiation Facility (ESRF). Susceptibility
data show an evolution between a spin glass state ({\itshape x}=0)
characterized by irreversibilities of field-cooled/zero field-cooled
curves below T$_{SG}$$\sim$ 22 K and a ferromagnetic state
({\itshape x}=0.2) indicated by a strong increase of magnetization
below T$_C$$\sim$ 58 K.

Magnetic diffraction patterns at ambient pressure were recorded on
the powder diffractometers G61 and G41 of the Laboratoire L\'eon
Brillouin (LLB) in the temperature range 1.4 K - 100 K. High
pressure neutron diffraction patterns were recorded on G61 in the
high pressure version\cite{Goncharenko04}. The neutron diffraction
patterns show the evolution of the magnetic order when going through
the critical threshold by substitution of Tb for La (Fig. 1a) and by
applied pressure (Fig. 1b). It is a clear evidence that magnetic
changes induced by chemical and applied pressures are equivalent.

In Tb$_2$Mo$_2$O$_7$ ({\itshape x}=0) short range correlations yield
diffuse maxima around {\itshape q}= 1 and 2 $\AA$$^{-1}$. An intense
signal at low {\itshape q} values shows the presence of
ferromagnetic correlations. For {\itshape x}=0.10, Lorentzian peaks
start to grow at the position of the diffuse maxima, revealing the
onset of mesoscopic magnetic order. For {\itshape x}=0.2, the low
{\itshape q} signal almost disappears and we clearly see magnetic
Bragg peaks showing long range magnetic order. When we apply
pressure on the {\itshape x}=0.2 sample, the intensity of the Bragg
peaks decreases and the ferromagnetic correlations and diffuse
maxima start to grow. The magnetic pattern for {\itshape x}=0.2 at
1.05 GPa (resp 3.7 GPa) is very similar to that for {\itshape
x}=0.10 (resp {\itshape x}=0) at ambient pressure.

In the range {\itshape q}= 0.5-2.5 $\AA$$^{-1}$, we analyzed the
magnetic correlations in Tb$_2$Mo$_2$O$_7$ by a short range magnetic
model in the same way as in Ref. \onlinecite{Greedan91}. A fit of
the diffuse magnetic intensity by the sum of radial correlation
functions was performed, giving information on spin-spin correlation
parameters $\gamma$ up to the fourth coordination shell ($\sim$ 7.3
$\AA$). The Tb-Tb correlations are F ($\gamma$$_{1,3,4}$$>$ 0),
while the Tb-Mo are AF ($\gamma$$_2$$<$ 0) in agreement with
previous results\cite{Greedan91}. The AF Mo-Mo correlations
responsible for the frustration in the SG state\cite{Gardner99}
cannot be detected, their contribution being about 50 times smaller
than the Tb-Tb ones due to the smaller Mo moment. The intense signal
below 0.5 $\AA$$^{-1}$ was not detected in previous patterns
measured in a higher {\itshape q} range\cite{Greedan91,Gaulin92}. It
cannot be accounted for by the short range magnetic model, even when
increasing the correlation range up to the seventh coordination
shell. We attribute this signal to ferromagnetic correlations
between Tb moments. We evaluate their lengthscale to about 20 $\AA$.

\begin{figure} [h]
\includegraphics* [width=\columnwidth] {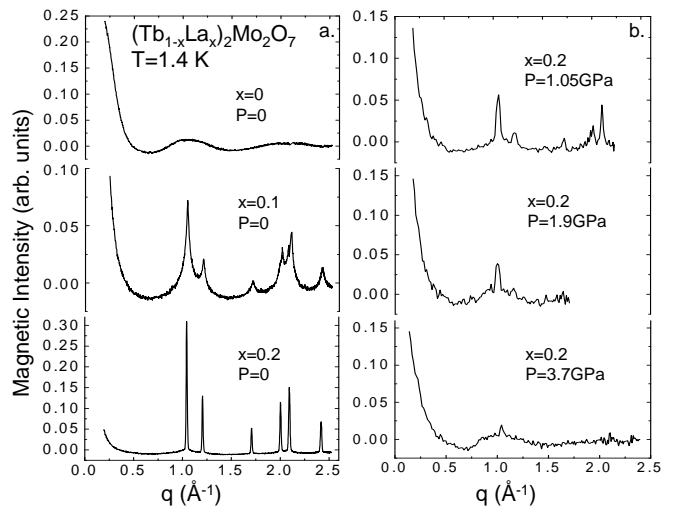}
\caption{Magnetic intensity of (Tb$_{1-x}$La$_x$)$_2$Mo$_2$O$_7$
at 1.4K versus the scattering vector
 {\itshape q}=4$\pi$sin$\theta$/$\Lambda$. The neutron wavelength is $\Lambda$=
 4.741 \AA.
 A spectrum in the paramagnetic phase (70 K) was
 subtracted and the magnetic intensity was scaled to the (222) nuclear peak intensity.
} \label{fig1.eps}
\end{figure}

The magnetic Bragg peaks observed for {\itshape x}=0.2 belong to the
face centered cubic lattice, showing that the magnetic structure is
derived from the chemical structure of {\itshape Fd$\overline{3}$m}
 symmetry by a propagation vector {\bf k}=0. The presence of two
magnetic peaks (200) and (220) forbidden in the pyrochlore structure
suggests a non-collinear F structure.

To study  the spin arrangement precisely, Rietveld refinements of
the magnetic diffraction patterns (Fig. 2) were performed with
FULLPROF \cite{Carvajal93}. The magnetic structure was solved by a
systematic search, using the program BasIreps\cite{BASIREPS} and
symmetry-representation analysis\cite{Izyumov91}. Since neither a
collinear F structure nor the {\bf k}=0 AF structure allowed by
{\itshape Fd$\overline{3}$m} symmetry were compatible with the
experimental data, we searched for a solution in the space group
{\itshape I4$_1$$/$amd}, the highest subgroup allowing F and AF
components simultaneously. The best refinement ({\itshape R$_B$}= 4
\%) is shown in Fig. 2a. In the ordered structure with {\bf k}=0,
the four tetrahedra of the unit cell are equivalent, for both the Tb
and Mo lattices. In a given Tb tetrahedron (inset Fig. 2a), the
Tb$^{3+}$ magnetic moments make an angle $\theta$$_t$= 11.6$^{o}$ at
1.4 K with the local $<$111$>$ anisotropy axes connecting the center
to the vertices. The components along these $<$111$>$ axes are
oriented in the "two in, two out" configuration of the local spin
ice structure \cite{Bramwell01}. The F component orders along a
$[$001$]$ axis. The Mo moments align close to a $[$001$]$ axis
(inset Fig. 2a), with a slight tilting by the angle $\theta$$_m$=
6.8$^{o}$ at 1.4 K towards the local $<$111$>$ axes. In this
structure, both Mo-Mo and Tb-Mo correlations are F, in contrast with
the spin glass described above. Interestingly for T$<$ 40 K a
diffuse intensity shows the onset of short range correlations. These
correlations coexist with the long range order, have the same
symmetry and their intensity increases with decreasing temperature.
A full description will be given in a future paper.

\begin{figure} [h]
\includegraphics* [width=\columnwidth] {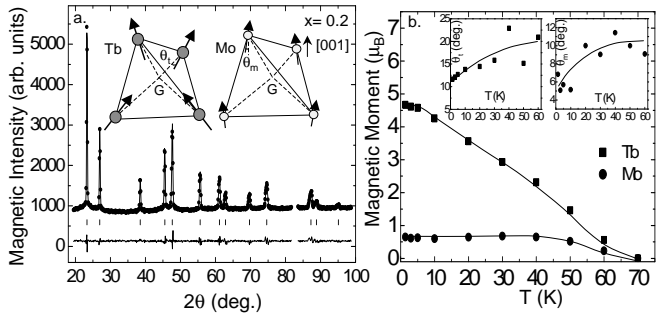}
\caption{a. Magnetic intensity in the {\itshape x}= 0.2 sample at
1.5 K versus the scattering angle 2$\theta$, with $\Lambda$= 2.426
\AA. A spectrum at 70 K was subtracted. Solid lines show the best
refinement ({\itshape R$_B$}= 4 \%) and the difference spectrum
(bottom). In the inset the magnetic structure of the Tb- and Mo -
tetrahedra. b. Ordered magnetic moments versus temperature. In inset
the angles $\theta$$_t$ and $\theta$$_m$ made by Tb- and Mo- moments
with the local anisotropy $<$111$>$ and the $[$001$]$ axes,
respectively. Solid lines are guides to the eye.} \label{fig2.eps}
\end{figure}

\begin{figure} [h]
\includegraphics* [width=\columnwidth] {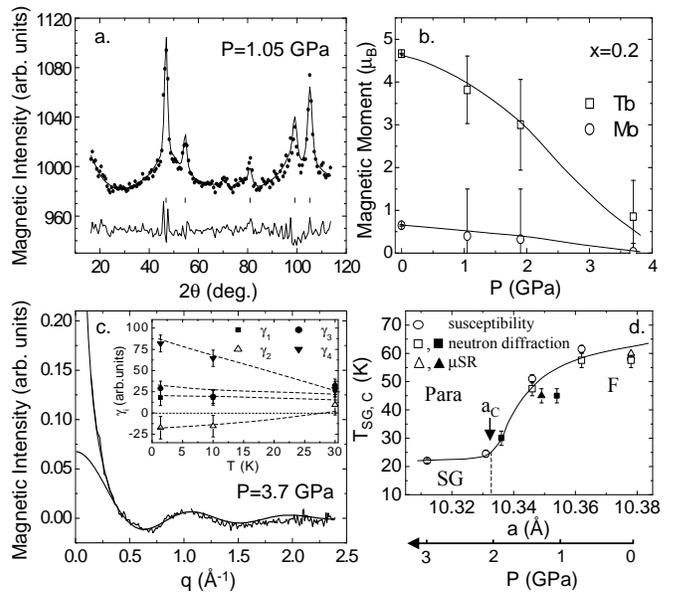}
\caption{a. Magnetic intensity in {\itshape x}= 0.2 sample at 1.4 K
and P=1.05 GPa, with $\Lambda$= 4.741 \AA. Solid lines show the best
refinement ({\itshape R$_B$}= 20 \%) and the difference spectrum; b.
Magnetic moment versus pressure; c. Magnetic intensity of {\itshape
x}= 0.2 at 1.4 K and P=3.7 GPa versus scattering vector. The fit is
made with the short range model (bottom line), including longer
range F correlations (upper line). In inset the temperature
dependence of the correlations coefficients; d. phase diagram for
(Tb$_{1-x}$La$_x$)$_2$Mo$_2$O$_7$ in the threshold region. Open
symbols are measured at ambient pressure for several {\itshape x}
contents. Full symbols are for {\itshape x}=0.2 under pressure,
taking into account ESRF data to determine {\itshape a}(P).}
\label{fig3.eps}
\end{figure}

Figure 2b shows the variation of Tb and Mo magnetic moments with
temperature. Below 40 K M$_{Mo}$ is almost T-independent, while
M$_{Tb}$ keeps increasing below T$_C$. The ordered moments at 1.4 K,
M$_{Tb}$= 4.66(1) $\mu_{\rm B}$ and M$_{Mo}$= 0.64(2) $\mu_{\rm B}$
are well reduced from the free ion values of 9 $\mu_{\rm B}$ and 2
$\mu_{\rm B}$, respectively. For Tb, this strong reduction could be
mainly explained by crystal field effects. As for Mo, it could arise
from a band effect and/or quantum fluctuations. The two tilting
angles slightly decrease with decreasing temperature (inset Fig.
2b).

The obtained non collinear structure both for Tb$^{3+}$ and
Mo$^{4+}$ originates from the uniaxial anisotropy of the Tb$^{3+}$
ion, which brings spin ice frustration to the ferromagnetic phase.
The ground state is determined by Mo-Mo F exchange interactions like
in Nd$_2$Mo$_2$O$_7$ \cite{Yasui01}, while in the "ordered spin ice"
Tb$_2$Sn$_2$O$_7$ with similar orientation of Tb$^{3+}$ moments
\cite{Mirebeau05} it results from F dipolar interactions between the
Tb$^{3+}$ ions.

Under pressure, the ordered moments ({\itshape x}= 0.2) decrease and
reorient (Fig. 3). For P=1.05 GPa, we observed the coexistence of
long and short range ordered phases of the same symmetry, yielding
Bragg peaks and a diffuse background respectively. The ordered Tb
moments keep the "two in, two out" spin configurations with a
different $\theta$$^\prime$$_t$= 28.3 $^{o}$, while the Mo ones turn
to a local spin ice structure, making an angle
$\theta$$^\prime$$_m$= 7.3 $^{o}$ with the local $<$111$>$ axes. At
3.7 GPa  the Bragg peak disappear. The near-neighbour correlations
parameters are similar to those of Tb$_2$Mo$_2$O$_7$ : F Tb-Tb
correlations ($\gamma$$_{1,3,4}$$>$ 0) and AF Tb-Mo correlations
($\gamma$$_2$$<$ 0). The ferromagnetic correlation length of 18
$\AA$ is also similar. The ordering temperature decreases with
increasing pressure.

$\mu$SR measurements (Fig. 4) shed a new light on the magnetic order
by probing the spin fluctuations and the static local field below
T$_C$. The recent availability of $\mu$SR under pressure allows us
to probe them on both sides of the threshold. We measured the
{\itshape x}=0.2 sample at ambient pressure on the GPS and GPD
instruments of the Paul Scherrer Institut (PSI) and under a pressure
of 1.3 GPa on GPD. The muon spin depolarization function P$_Z$(t)
for several spectra recorded at ambient pressure is shown in Fig 4a.

$\mu$SR spectra above T$_C$ were best fitted with a stretched
exponential function P$_{Z}$(t)=exp(-$\lambda$t)$^\beta$. Below
T$_C$, P$_{Z}$(t) was fitted by the function P$_Z$ (t)=
[exp(-$\lambda$$_Z$t)$^\beta$ + 2 exp(-$\lambda$$_T$t)
cos($\gamma$$_{\mu}$$<$B$_{loc}$$>$t)]/3, expected for the the
magnetically ordered state of a powder sample\cite{Dalmas04}. The
first term of function below corresponds to the depolarization by
spin fluctuations perpendicular to the direction of the muon spin,
whereas the second term reflects the precession of the muon spin in
the average local field $<$B$_{loc}$$>$ at the muon site. The
transverse relaxation rate $\lambda$$_{T}$ could have both static
and dynamical character. Both expressions of P$_{Z}$(t) are expected
to merge in the high temperature limit, when the dynamics of
Tb$^{3+}$ and Mo$^{4+}$ moments is fast, yielding
$\lambda$$_{Z}$=$\lambda$$_{T}$, $<$B$_{loc}$$>$=0 and $\beta$=1.
One should also mention that a fit with a dynamical Kubo-Toyabe
(DKT) depolarization function is possible for a small temperature
range below T$_C$, but the function below gives better results at
low temperature and close to T$_C$.

\begin{figure} [h]
\includegraphics* [width=\columnwidth] {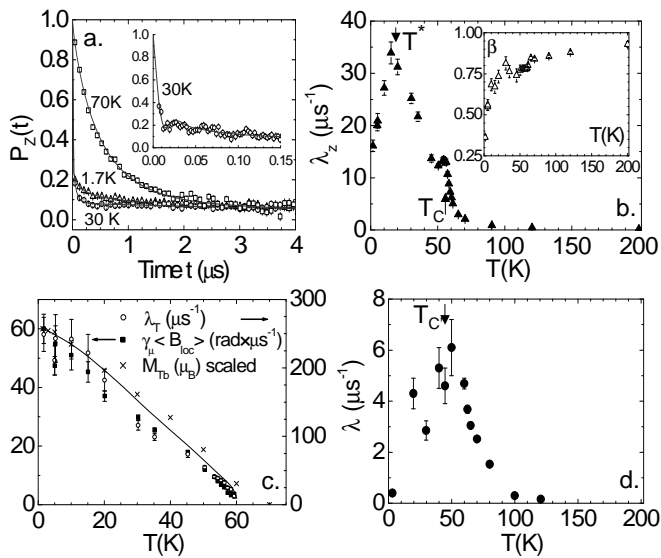}
\caption{$\mu$SR results for {\itshape x}=0.2. a. Muon
depolarization function P$_Z$(t) at ambient pressure for several
temperatures; b. Temperature dependence of : $\lambda$$_{Z}$ and
$\beta$ at ambient pressure; c. $<$B$_{loc}$$>$,  $\lambda$$_{T}$,
and M$_{Tb}$ (scaled) at ambient pressure;
 d. $\lambda$ at P=1.3 GPa .} \label{fig4.eps}
\end{figure}

The longitudinal relaxation rate rate $\lambda$$_{Z}$ clearly shows
a cusp at T$_C$ $\sim$ 60 K. Surprisingly, with decreasing T
further, $\lambda$$_{Z}$ does not keep decreasing but starts to
increase again below about 50 K, showing a broad maximum at
T$^\star$ $\sim$ 25 K. The behavior around T$^\star$ is clearly
non-critical as $\lambda$$_{Z}$(T) does not tend to diverge. The
transverse relaxation rate $\lambda$$_{T}$ is about 10 times larger
than $\lambda$$_{Z}$(T). It smoothly increases below T$_C$ and
scales with the average local field $<$B$_{loc}$$>$ as temperature
varies. This leads us to assign $\lambda$$_{T}$ mainly to the width
of the distribution of local fields. Interestingly both quantities
also scale with the ordered moment M$_{Tb}$(T) measured by neutron
diffraction. It suggests that the local field seen by the muon
mostly comes from the Tb$^{3+}$ ions with much larger moments,
although more localized, than the Mo$^{4+}$ ones. Previous $\mu$SR
data in spin glasses \cite{Dunsiger96} support this interpretation,
showing that the static internal field is about 10 times larger in
Tb$_2$Mo$_2$O$_7$ (0.7 T) than in Y$_2$Mo$_2$O$_7$ (0.066 T).

Under pressure and for temperatures below T$_C$ it was difficult to
extract any information from the $\mu$SR spectra at small times (the
2/3 term) , due both to the large background of the pressure-cell
and to the fast depolarization of the 2/3 term. Therefore, below
T$_C$, we fitted only the 1/3 term, with an exponential
depolarization function, skipping the first 0.2 $\mu$s of the
$\mu$SR spectra. There is no anomaly corresponding to T$^\star$,
suggesting that the temperatures T$_C$ and T$^\star$ merge with
increasing pressure, as the sample enters in the spin glass-like
state. The fitted values of $\lambda$ are clearly lower under
applied pressure.

The second transition at T$^\star$ seen by $\mu$SR recalls
observations in disordered frustrated
 ferromagnets called reentrant spin glasses (RSG)
 \cite{Mirebeau97,Ryan00}. In RSG's, AF interaction compete with dominant F interactions,
 leading to spin glass-like anomalies below T$_C$.
The second transition was assigned to the freezing of spin
components transverse to the domain magnetization without breaking
off long range ferromagnetic order, as predicted by mean field
theory\cite{Gabay81}. Here the physical meaning of T$^\star$ should
be more complex since the long range magnetic order 
involves canted moments. We notice that there is no change in this
order at T$^\star$, as could be seen by an anomaly in the ordered
moments or canting angles. We suggest a freezing of short range
correlated moments below T$^\star$ since we observe diffuse
scattering down to the lowest temperature. We could check it in
future by inelastic neutron scattering.

In conclusion, we studied for the first time the microscopic
magnetic properties at the F-SG threshold. Using pressure, we prove
the dominant role of Mo-Mo distances and clarify the role of R
magnetism. We also observe a new dynamical transition on the
ferromagnetic side of the threshold.

We thank A. Amato and U. Zimmermann for $\mu$SR measurements on GPS
and GPD (PSI), W. Crichton for X ray measurements on ID27 (ESRF), F.
Bour\'ee and G. Andr\'{e} for neutron measurements on 3T2 and G41
(LLB), A. Forget and D. Colson for the sample preparation (SPEC).

\end{document}